# Revealing the origins of shear band activity and boundary strengthening in polygrain-like architected materials

*Chen Liu, Jedsada Lertthanasarn &Minh-Son Pham*

Department of Materials, Imperial College London, Exhibition Road, South Kensington, London, SW7 2AZ, United Kingdom

Keywords: Meta-crystal lattice, architected materials, shear band, grain boundary strengthening

**Highlights**

- Shearing activities in FCC-architected materials were identified.
- Significant strengthening and stabilisation induced polygrain-like boundaries.
- Enabling great control of shearing behaviour in architected materials.
- The type and coherency of boundaries are influential to the strengthening.

**Abstract**

A recent report on successful employment of the grain boundary strengthening to design extraordinarily damage-tolerant architected materials (i.e. meta-crystals) necessitates fundamental studies to understand the underlying mechanisms responsible for the toughening and high performance of meta-crystals. Such understanding will enable greater confidence and control in developing high performing and smart architected materials. In this study, buckling of lattice struts in single crystal-like meta-crystals was firstly analysed to reveal its role in shear band activities. Shear band systems of singly oriented meta-crystals were also identified to provide a solid basis for predicting and controlling the shearing behaviour in polygrain-like meta-crystals. The boundary-induced strengthening effects in meta-crystals was found to relate to the boundary type and coherency as they govern the transmission of shear bands across meta-grain boundaries. The obtained insights in this study provide crucial knowledge in developing high strength architected materials with great capacity in controlling the mechanical strength and damage path.

**Graphical abstract**

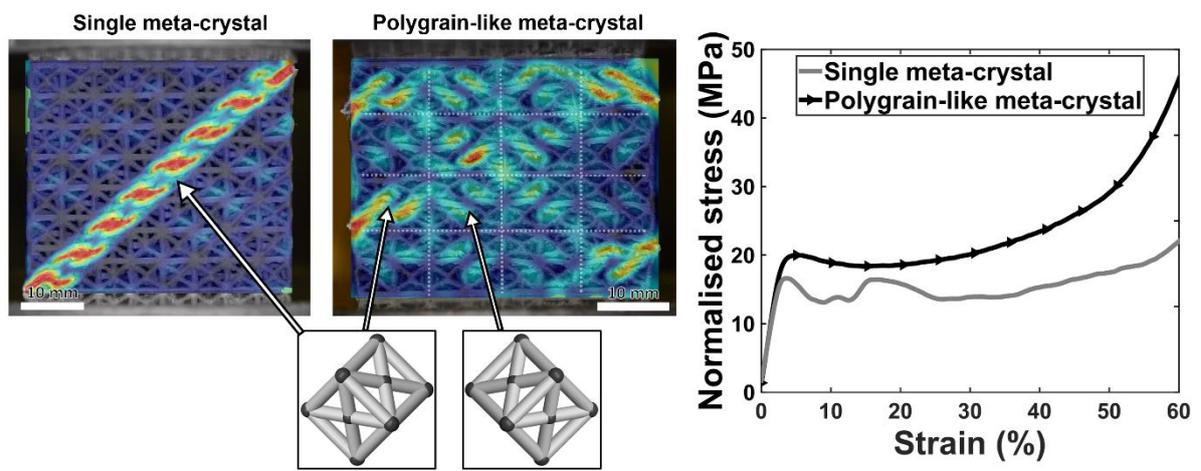

# 1. Introduction

Architected materials built on ordered arrangements of internal structures are lightweight and excellent at bearing load and absorbing impact energy, enabling their wide applications in the fields of aerospace, automobiles, medical devices, packaging and infrastructures [1–4]. Lattice materials formed by periodic arrangement of unit cells consisting of a regular network of struts are one particular type of architected materials [5,6]. Extensive and in-depth works have been carried out with focusing on optimising the elastic, yielding and toughness properties of singly-orientated lattice materials among various unit cells types, such as octet-truss [7,8], body centred cubic [9–12], gyroid [13] and auxetic [14]. Reviewing different singly-orientated lattice materials, one of the common issues when bearing load is the undesirable severe drop in strength after plastic yield (post-yield collapse) due to the formation of dominant and fast propagating shear bands throughout the whole structure, in particular for lattice materials made of elasto-plastic materials or containing process defects formed in manufacturing [15–18]. As a consequence, the strength and energy absorption capacity of lattice materials is substantially reduced over a wide deformation range. With the aim of designing high performance architected materials, it is, therefore, desirable to derive the method of manipulating the shearing behaviour of lattice materials to minimise such shearing related weakening effects.

A transformative approach of designing new extraordinary damage tolerant architected materials (coined meta-crystals) was recently proposed to successfully translate the metallurgical grain boundary hardening mechanisms in crystalline materials to eliminate the post-yield collapse in lattice materials by designing polygrain-like structures [19]. The study reveals that the strength and stability of architected materials can be dramatically increased thanks to the significantly shortening and impeding effects of meta-grain boundaries on the dominant shear band. Such boundary impeding effects on shear bands are attributed to not only the change of lattice orientations across a boundary [19], but also the high dependence of shearing behaviour on lattice orientations, which relationship is crucial and has not been reported. Therefore, identification of the shear band activity in singly-orientated lattice structure will help to understand and predict the spatial distribution of shear bands in polygrain-like meta-crystals, hence improving the control of damage propagation in meta-crystals.

Moreover, our previous study interestingly showed that the yield strength of meta-crystals increases with a reduction in the size of meta-grains, and such a correlation can be described by an empirical model that is similar to the Hall-Petch relationship found for polycrystalline metal [20–24]. Although the strengthening effects via reducing meta-grain size was clearly presented and attributed to the role of meta-grain boundaries in temporarily stopping the shear band propagation [19], underlying mechanisms responsible for the observed strengthening have not been discussed. The theories in physical metallurgy indicate that the coherency at grain boundary influences the strength of boundary and the slip transmission across boundaries, hence significantly contributing to the grain boundary hardening [25,26]. The effect of boundary coherency is also associated with the types of boundaries, such as tilt or twist boundaries, leading to different degrees of strengthening of the boundary type in alloys [23]. Therefore, this study will investigate the role of boundary coherency on the transmission of shear bands across meta-grains, providing insights into the boundary strengthening in architected materials. In particular, the study examines if there was a correlation between the type of boundary and strength of meta-crystals.

In this study, shear band systems in differently orientated lattice structures inspired by a face-centred-cubic (FCC) crystal were studied and identified. In particular, buckling analysis was done to explain and predict the shearing activities in single-grain-like meta-crystals. Subsequently, meta-crystals (mimicking polygrains) containing numbers of meta-grains ranging from one to sixty-four were designed and tested to study the shear band propagation in orientated lattice meso-structures and transmission across meta-grain boundaries. The role of boundary and its coherency on shearing activities, yield strength and hardening are revealed and discussed.

## 2. Methods
### 2.1. Lattice design

A cubic unit cell with dimension of $5 \times 5 \times 5 mm^3$ (shown in Fig. 1a) inspired by FCC atomic lattice was designed using nTopology Element. A global orthogonal coordinate system was right-handed with global Z-axis being parallel to the build direction of fabrication and also the loading direction of compression test. To study the orientation dependence, architected materials containing singly-orientated unit cells of architected lattices (i.e. equivalent to single crystals) were designed. The first single meta-crystal was generated in two rotation steps as follows: the three ⟨1 0 0⟩ directions of the unit cell were initially aligned parallel to the global X-, Y- and Z-axes (Fig. 1b). The unit cell was then rotated 45° clockwise about the global X-axis (Fig. 1c) and followed by 30° clockwise about the global Y-axis (Fig. 1d). The whole lattice structure was then constructed by tessellating the rotated unit cell to infill a global cube with a side length of $40mm$. Similarly, two additional lattices were created following the same procedure for the first rotation, however with a second rotation of 45° and 60° about the global Y-axis (Fig. 1e and f). In total, three singly orientated meta-crystals with their [1 0 0] inclined at different angles to the loading direction were created, referred to as FCC30, FCC45 and FCC60 depending on the unit-cell rotation about the Y-axis.

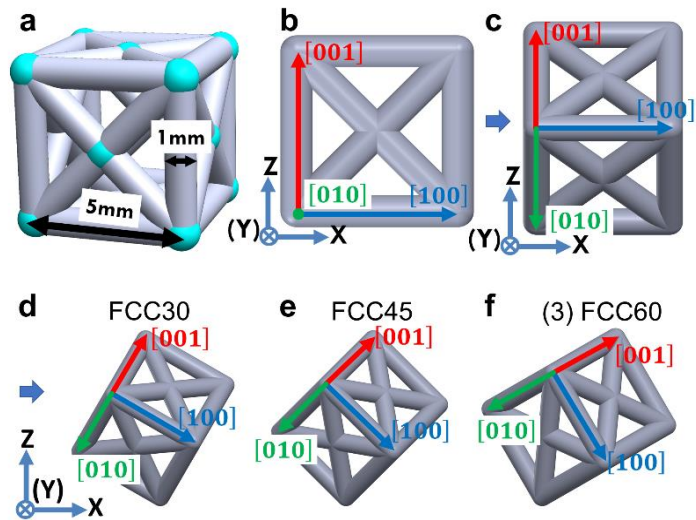

Fig. 1. (a) FCC unit cell, (b-f) the rotation sequence of unit cell: (b) initial unit cell, (c) unit cell is firstly rotated 45° about the X-axis, (d-f) unit cells after second rotation about the Y-axis: (d) FCC30, (e) FCC45 and (f) FCC60.

The same design approach as presented in our previous study was used to create polygrain-like architected materials (i.e. meta-crystals)[19]. The global lattice structure with dimensions of $40 \times 40 \times 40 mm^3$ was divided into domains (called meta-grains) of which the lattice orientation was tailored to be different to that of its adjacent meta-grains. For example, for a meta-crystal containing two twinned meta-grains (Fig. 2a), in the left meta-grain (Fig. 2b), the unit cell was rotated following the same rotation sequence as that of FCC45 (shown in Fig. 1e). Whereas, in the right meta-grain, the unit cell was rotated by the same magnitude in the same sequence, but in the anti-clockwise rotation direction (Fig. 2b). Note that the rotation of the internal structure in each meta-grain leads to incomplete unit cells in the vicinity of the meta-grain boundaries and therefore generates open struts that would weaken a meta-crystal at meta-grain boundaries, 2D square planar frame with unit cell side length of $5mm$ was inserted at the boundary between two meta-grains. This allows for the open struts to connect to the boundary and maintain the connectivity. One such frame structure at a boundary is highlighted in Fig. 2c. Meta-crystals containing 4, 8, 16, 32 and 64 meta-grains were constructed with boundaries as highlighted (Fig. 2d-h). It is noted that, with the consideration of the size of FCC unit cell (Fig. 1b), dividing meta-crystal into 64 meta-grains is nearly the finest strategy to enable one meta-grain contain one complete FCC unit cell. The size of a meta-grain was calculated as the diameter of a sphere that has the same volume to the meta-grain.

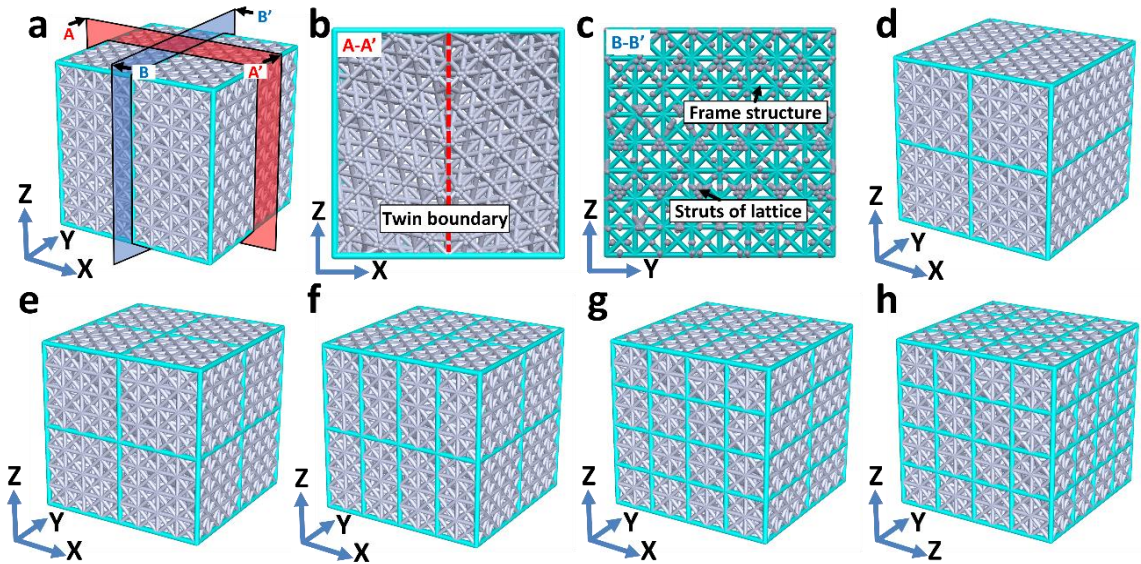

Fig. 2. (a) Meta-crystal containing 2 meta-grains, (b) A-A' cross-sectional view highlighting the twin boundary of the meta-crystal shown in (a), (c) B-B' section of (a) showing the 2D frame (highlighted by light green colour) which acts as the boundary (note: lattice struts connected to the boundary were shown in grey), (d-h) meta-crystals containing 4, 8, 16, 32 and 64 meta-grains, respectively.

## 2.2. Experiments and post-analysis

The designed meta-crystal models were fabricated using Fused Deposition Modelling (FDM) 3D printer (Ultimaker 2) with polylactic acid (PLA) filament obtained from RS components. Some main printing parameters were given as followings: layer thickness of $0.1mm$, printing speed of $30mm \cdot s^{-1}$, nozzle travel speed of $120mm \cdot s^{-1}$, nozzle diameter of $0.4mm$, printing temperature of $210°C$ and build plate temperature of $60°C$. The build direction was parallel to the Z-axis shown in Fig. 2.

Compression tests were carried out by a $100kN$ Zwick testing machine at a strain rate of $0.001s^{-1}$ at room temperature. Lubricant (dry molybdenum disulphide) was used to minimise the effect of friction between lattices and compression plates. The loading direction was parallel to the Z-axis shown in Fig. 2. Nominal engineering stresses were calculated by dividing the recorded forces by the nominal cross-section area ($41 \times 41 mm^2$). Engineering strains were derived by dividing the change in the length along the Z direction by the initial length $41mm$. Images of the front face of the meta-crystal (parallel to global X-Z plane shown in Fig. 2) were taken at one second intervals by a Nikon D7100 camera with 200mm Nikkor macro lenses. The series of captured images during the deformation were analysed by Digital Image Correlation (DIC) via a commercial software DaVis with following settings: image size of $6000 \times 4000\ pixels$, subset size of $101 \times 101\ pixels$ and a step size of $25\ pixels$.

The mechanical properties of the base material were obtained by uniaxial tensile test on specimens designed according to ISO 527 standard. Specimens were fabricated by FDM using the same printing parameters as the meta-crystals. Average elastic modulus $E_s$ and yield stress $\sigma_{y\_s}$ of the base material were calculated from three tensile stress-strain curves, giving $E_s = 1190.4 MPa$ and $\sigma_{y\_s} = 47.6 MPa$.

### 2.3. Buckling force analyses

To understand the deformation behaviours of lattices under loading, we performed force analyses on the struts of FCC unit cells orientated to different directions with respect to the loading direction. Fig. 3a demonstrates one example of the relationship between unit cell orientation and loading direction that is parallel to the Z-axis. The directions of individual struts were defined with respect to the unit cell's local coordinate system (shown in Fig. 3b). To analyse a certain strut, the compressive force $F$ was assumed to be parallel to the Z-axis which is subsequently resolved into two orthogonal forces: $F_{ra}$ (along the axis of strut) and $F_{rp}$ (perpendicular to $F_{ra}$ and within the plane formed by the axis of strut and global Z-axis) (shown in Fig. 3c). Since the buckling is the main failure mode of slender struts, critical buckling force $F_{cr}$ is calculated through Eqn.1 from Euler's buckling criteria [27], where $E_s$ is the elastic modulus of base material, $I$ is the second moment of area, $K$ is the strut effective length factor that is dependent on the end fix of struts ($K$ should be almost the same for every struts due to the same end fix), and $L$ is the length of strut. For inclined strut shown in Fig. 3c, the critical buckling force was along global Z-axis $F = F_{cr}/sin\theta$. Taking that $I = AR_g^2$, where $A$ and $R_g$ are the area and radius of gyration of the cross section of struts, critical buckling stress $\sigma_{cr}$ can be obtained by Eqn. 2. In addition, the buckling tendency of strut can be determined by the slenderness ratio $S = KL/R_g$. Because the buckling causes the shear band, resulting in the plastic deformation of the lattice material, the critical buckling slenderness ratio of the used PLA base material can be calculated by assuming the critical buckling stress equal to the yield stress of the base material (Eqn. 3), giving $S_c = 15.7$. Thus, strut with slenderness ratio larger than 15.7 are more susceptible to buckling.

$$F_{cr} = \frac{\pi^2 E_s I}{(KL)^2} \quad (1)$$

$$\sigma_{cr} = \frac{F_{cr}}{A} = \frac{\pi^2 E_s R_g^2}{(KL)^2} \quad (2)$$

$$S_c = \pi \sqrt{\frac{E_s}{\sigma_y}} \quad (3)$$

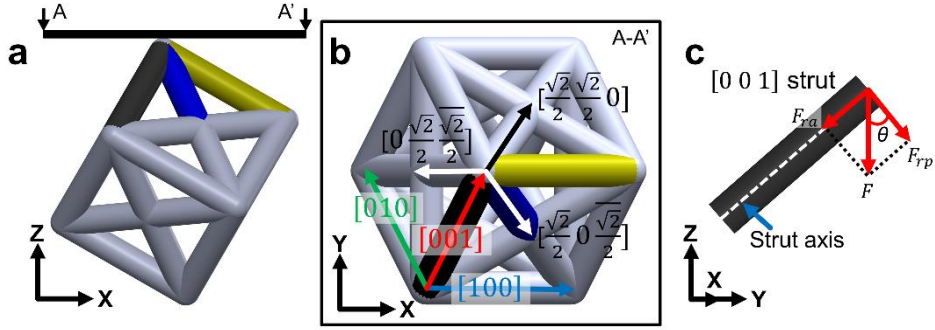

Fig. 3. Force analyses on FCC unit cell, (a) orientated unit cell with respect to loading direction parallel to the global Z-axis, (b) identification of individual struts, where [1 0 0], [0 1 0] and [0 0 1] struts belong to a group of ⟨1 0 0⟩ struts, while $\left[\frac{\sqrt{2}}{2} \frac{\sqrt{2}}{2} 0\right]$, $\left[\frac{\sqrt{2}}{2} 0 \frac{\overline{\sqrt{2}}}{2}\right]$ and $\left[0 \frac{\sqrt{2}}{2} \frac{\overline{\sqrt{2}}}{2}\right]$ struts belong to a group of ⟨$\frac{\sqrt{2}}{2} \frac{\sqrt{2}}{2} 0$⟩ struts, (c) force resolution applied on strut.

### 2.4. Finite Element Analysis (FEA)

Finite Element Analysis (FEA) was carried out to study the deformation behaviour of internal architected struts in single meta-crystals using Abaqus/Standard to assist DIC analyses. Two simplified lattice models, namely FCC45 and FCC60, are designed for FEA with dimensions of $25 \times 15 \times 7.42 mm^3$, in which FCC unit cell with side length of $5mm$ was used (Fig. 4). Lattice frame structure for boundary is comprised by rectangular units with dimensions of $3.57 \times 3.71 mm^2$ (Fig. 4a). Designed lattice models were imported into Abaqus/CAE and meshed using B31 beam elements with seeds size of $0.5mm$, where the elemental beam orientation was assigned to be along the direction of strut. Materials properties for lattice in FEA were obtained from the uniaxial tensile tests of 3D printed specimen (see Section 2.2). Two rigid plates, namely top and bottom plate, were created and meshed by R3D4 rigid elements with seeds size of $2mm$. Tie constraints were applied for top and bottom lattice nodes that are in contact with the top and bottom plates, respectively (Fig. 4d). For the boundary conditions, the bottom plate was fixed while the top plate moved down for compression at a strain rate of $0.001s^{-1}$.

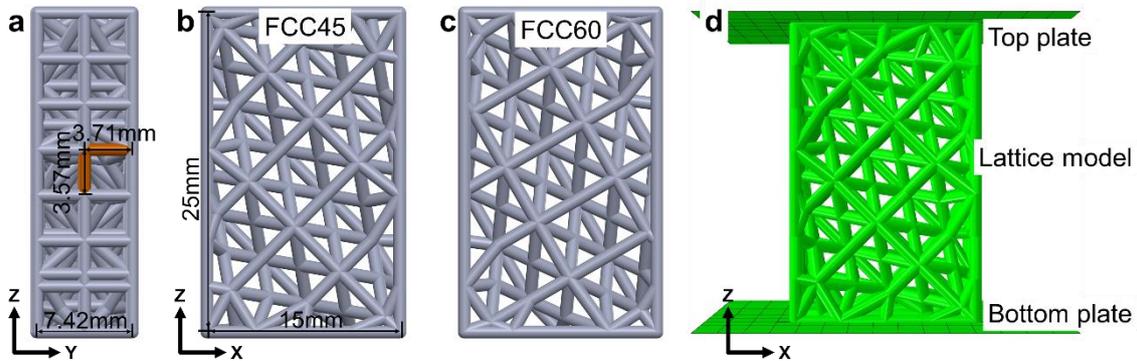

Fig. 4. FEA lattice models designed for FEA simulation, (a) sideview (Z-Y plane) shows boundary frame and dimensions, (b) FCC45, (c)FCC60, (d) Boundary conditions for FEA.

## 3. Results and discussion

### 3.1. Shearing behaviour in singly-orientated meta-crystals

The stress-strain responses of single meta-crystals with different orientations regarding to loading directions are shown in Fig. 5a. A clear drop in strength after yielding (occurring at strain of around from 3% to 10% strain) is observed in all three lattices. This is due to the collapse of struts along a specific direction, forming parallel shear bands shown in Fig. 5b-d in consistent with previous reports on singly orientated lattice materials [15–18]. The relationship between shear bands and lattice orientations were further revealed in DIC analyses. Frame insets of orientated unit cells indicate the shear direction in relation to the orientation of unit cell of the infilled structure (Fig. 5b-d). Similar to the determination of slip systems (which consist of slip direction and slip plane) in crystalline alloys, shearing systems in meta-crystals can be defined by the direction and plane of shearing. The shear systems in FCC meta-crystals designed in this study were identified to be $\{0\ 2\ 2\}\langle 1\ 0\ 0\rangle$ in FCC30 and $\{2\ 0\ 0\}\langle 0\ 1\ 1\rangle$ in FCC45 and FCC60. The identification of shear planes for different lattice types showed that the $\{0\ 2\ 2\}$ planes were observed in BCC [17], F2CCz [18] and BCCZ [28] architected materials while shear planes $\{0\ 0\ 2\}$ were reported for simple cubic [29]. However, these previous studies did not report the shear directions and were only confined to a single loading direction with respect to the lattice orientation, limiting the understanding of shear band activities in multi-orientated lattices. The obtained results of shearing behaviour show that there are at least two different shearing systems for the studied FCC meta-crystals. This identification is important for accurately controlling the load transfer and damage propagation along specific directions in polygrain-like meta-crystals.

To acquire in-depth understanding on the activation of these shearing systems in the meta-crystals, critical force for buckling of individual struts (Fig. 3b, struts are identified according to their directions) in each orientated lattice was analysed. For the unit cell used in this study, constructed struts can be classified into two groups by their geometries: $\langle 1\ 0\ 0 \rangle$ struts with slenderness ratio $S = 20$ which is larger than the critical ratio $S_c = 15.7$ (Eqn. 3), and $\langle \frac{\sqrt{2}}{2}\ \frac{\sqrt{2}}{2}\ 0 \rangle$ struts with $S = 14.14$ which is smaller than $S_c$. Thus, $\langle 1\ 0\ 0 \rangle$ struts should tend to buckle first, leading to shear band formation along directions on planes perpendicular to $\langle 1\ 0\ 0 \rangle$ in the FCC-inspired meta-crystals. Moreover, critical buckling forces along global Z-axis for $[0\ 1\ 0]$ and $[1\ 0\ 0]$ struts (Fig. 5e) indicate that $[0\ 1\ 0]$ strut buckled first in FCC30 while buckling should first occur for the $[1\ 0\ 0]$ struts in FCC45 and FCC60. Such buckling sequence explains the formations of shear band systems as experimentally observed by DIC analyses (Fig. 5b-d).

In contrast to very well-defined shear activities in FCC30 (Fig. 5b) and FCC45 (Fig. 5c), the shear bands in FCC60 appear more complex, causing difficulties in the shear band identification. In addition to the dominant $\{2\ 0\ 0\}\langle 0\ 1\ 1\rangle$ shear band, some fine and localised shear bands were observed in FCC60 lattice (Fig. 5d). This shearing behaviour was identified to be $\{0\ 2\ 2\}\langle 1\ 0\ 0\rangle$, which might also be another shear band system for FCC60. To clearly identify the shear band system in FCC60, FEA was used to help understand the deformation behaviour of struts in this meta-crystal. FEA simulation set-up was first validated against the experimental observation for FCC45 (Fig. 5f vs Fig. 5c). The simulation shows that $[1\ 0\ 0]$, $\left[\frac{\sqrt{2}}{2}\ \frac{\sqrt{2}}{2}\ 0\right]$ and $\left[\frac{\sqrt{2}}{2}\ 0\ \frac{\sqrt{2}}{2}\right]$ struts were highly rotated, forming a localisation of deformation along $\{2\ 0\ 0\}\langle 0\ 1\ 1\rangle$ shear band that is consistent with the shear band observation from DIC analyses (Fig. 5c), confirming the validity of simulation. The FCC60 was then simulated by using the FEA parameters identified for FCC45. Fig. 5g confirms that $\{2\ 0\ 0\}\langle 0\ 1\ 1\rangle$ is the main shear band system in FCC60 while $\{0\ 2\ 2\}\langle 1\ 0\ 0\rangle$ was not seen in the FEA simulation. In addition, DIC analysis (Fig. 5d) shows that $\{0\ 2\ 2\}\langle 1\ 0\ 0\rangle$ was only observed locally near collapsed struts of $\{2\ 0\ 0\}\langle 0\ 1\ 1\rangle$. Similarly, Fig. 5b shows minute and short localisation along a different direction to the $\{0\ 2\ 2\}\langle 1\ 0\ 0\rangle$ in FCC30. Therefore, the shortened localisation in

{0 2 2}⟨1 0 0⟩ observed in FCC60 was likely due to the local stress re-distribution once a {2 0 0}⟨0 1 1⟩ shear band formed. The experimental and simulation analyses confirm that {0 2 2}⟨1 0 0⟩ was not a shear band system, but {2 0 0}⟨0 1 1⟩ is the only shear band system for the FCC60.

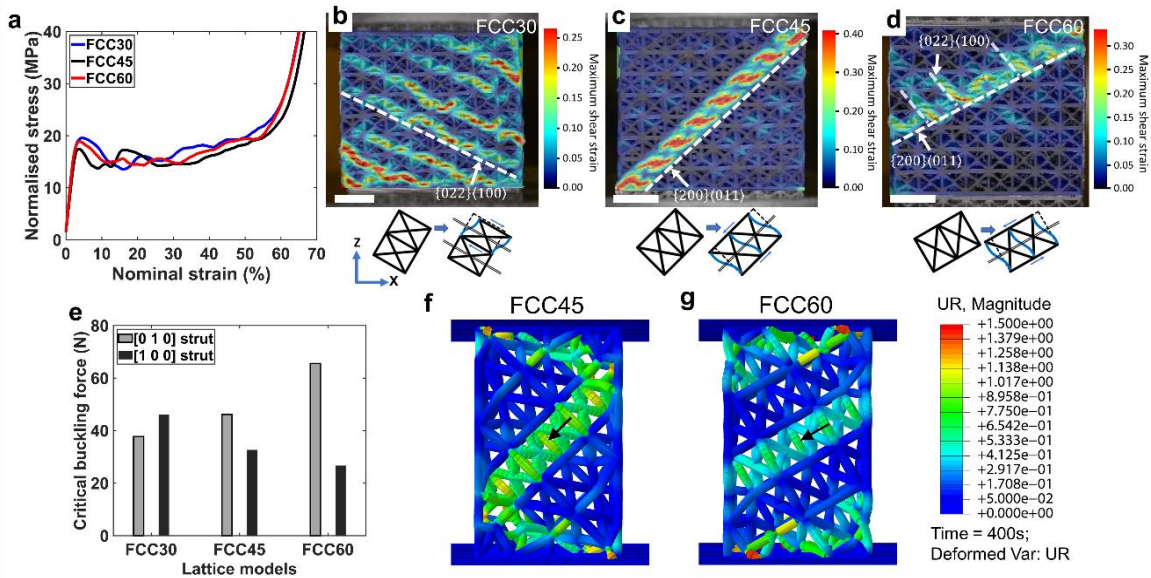

Fig. 5. Deformation in singly orientated materials: FCC30, FCC45 and FCC60, (a) Stress-strain curves, where stress was normalised by respective relative densities, (b-d) Localised deformation behaviour revealed by DIC analyses – cubic frame insets were used to show the lattice orientation and shearing direction, (e) critical buckling force calculated for unit *K* along global Z-axis, (f-g) FEA simulations at 40% strain for FCC45 and FCC60, respectively, showing rotational displacement at deformed nodes.

### 3.2. Meta-grain boundary strengthening

The effect of meta-grain boundary is initially revealed by comparing two meta-crystals containing one and two meta-grains individually, i.e. single meta-crystal and bi-meta-crystal, respectively (Fig. 6). Stress-strain curves of both meta-crystals are shown in Fig. 6a. As the unit cell in each meta-grain were aligned to the same angle to the global maximum shear stress, each meta-grain should have the same yield strength due to buckling. Therefore, the buckling-induced yield strength of the single and bi-meta-crystals were expected to be the same. Interestingly, however, the measured yield strength and the firstly reached peak strength of the bi-meta-crystal was slightly higher than that of the single meta-crystal, highlighting the role of the boundary separating the two meta-grains. In addition, although both meta-crystals show softening behaviour after reaching the yield stress, the post-yield collapse in the bi-meta-crystal was much less severe. The hardening rate (measured by the variation of flow stress $\Delta\sigma$ over a strain range $\Delta\varepsilon = 0.01$) was calculated to quantify the improvement in stabilising the deformation behaviour of meta-crystals (Fig. 6b). Positive rates represent the hardening while negative rates represent softening. The absolute value of the largest softening rate ($34.6 MPa$) for the single meta-crystal wass almost double than that ($18.3 MPa$) of the bi-meta-crystal, suggesting the stress drop was much less severe in the bi-meta-crystal. In addition, Fig. 6b shows that the rate fluctuated drastically for the single meta-crystal while the fluctuation was much smaller for the bi-meta-crystal, i.e. the meta-crystal containing two meta-grains behaves in a much more stable manner than the single meta-crystal due to the presence of the boundary between the two meta-grains. Fig. 6c and d shows that a single and straight shear band in the single meta-crystal was formed along the maximum shear stress, resulting in an abrupt drop in measured stress (Fig. 6a) and the hardening rate (Fig. 6b). In contrast, two diffused shear bands were formed and aligned in two different directions which were symmetrical about the twin boundary in the bi-meta-crystal (Fig. 6e and f). Shear bands in each meta-grain were identified to be of the $\{2\ 0\ 0\}\langle 0\ 1\ 1\rangle$ system in consistent with observation in the single meta-crystal FCC45 presented in Section 3.1. The two symmetrical shear bands were much shortened (almost a half of that in the single meta-crystal) and confined by the twin boundary introduced between the two meta-grains, confirming the boundary was able to stop and deflect shear bands, minimising the stress drop after yielding. In addition to the shortened shear bands, the deformation in the twinned bi-meta-grains were more homogeneous and uniform with a much lower degree of localisation than in the single meta-crystals (Fig. 6f vs d) – Note that the scale bar of the maximum shear strain in the bi-meta-crystal is smaller than that in the single meta-crystal. In other words, the boundary significantly stabilised the deformation after yielding in the bi-meta-crystals by inhibiting the propagation of shear bands and diffusing the localised deformation.

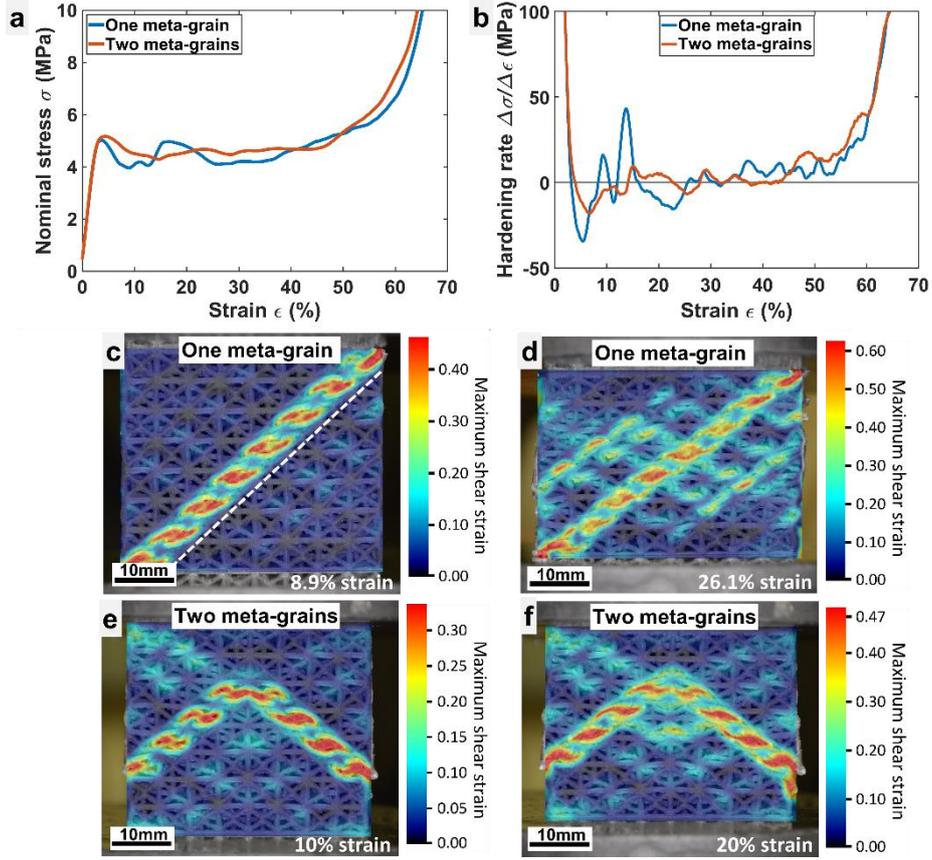

Fig. 6. Comparison between single and bi-meta-crystals, (a) stress-strain curves, (b) hardening rate, (c-f) deformation behaviours of each lattice at given strains after DIC analyses.

The strengthening effect of meta-grain boundary on meta-crystals is further highlighted by reducing the meta-grain size. Fig. 7a shows the normalised stress-strain curves of meta-crystals with different numbers of meta-grains (including the twinned bi-meta-crystal). Reducing the meta-grain size results in a general increase in strength, which is consistent with our previous study [19] (note that the stress-strain in this current study includes data points of smaller meta-grains and was normalised by the relative density of architected materials). In addition, it was reported that a relationship between the nominal yield strength ($\sigma_Y$ at 0.2% strain offset) and the size of meta-grain that can be described by a Hall-Petch-like relationship, i.e. $\sigma_Y = \sigma_0 + \frac{k}{\sqrt{d}}$ (where $d$ is the size of meta-grains, $\sigma_0$ is frictional stress and $k$ is a material constant). A least square linear fitting gives relationships: $\sigma_{Y\_L} = 12.49 + \frac{14.46}{\sqrt{d}}$ (Fig. 7b), reaffirming the applicability of Hall-Petch relationship in meta-crystals with elimination of the effects come from relative density.

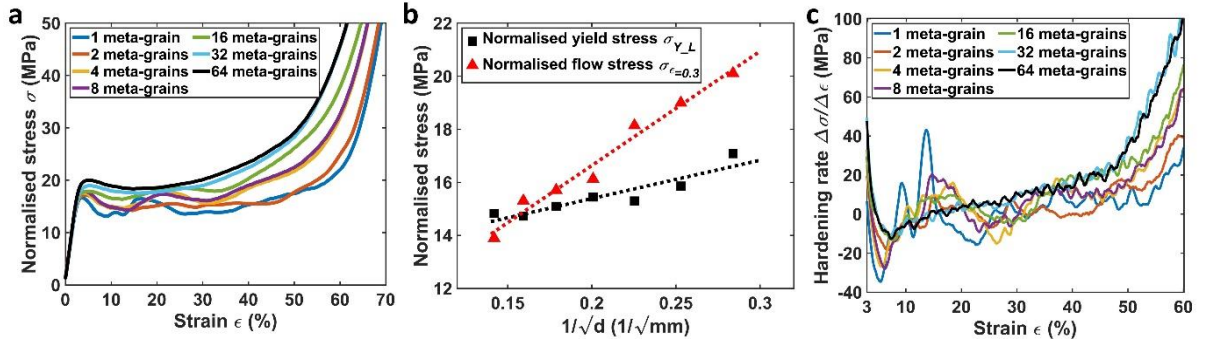

Fig. 7. (a) Stress-strain curves of lattices with different number of meta-grains (note: stresses are normalised by the relative density of individual meta-crystal), (b) relationship between normalised yield stress $\sigma_{Y\_L}$, flow stress $\sigma_{\varepsilon=0.3}$ at $\varepsilon = 0.3$ and meta-grain size, (c) hardening rates.

The reduction in the size of meta-grains also improved post-yield strength and increasingly stabilised the flow stress of meta-crystals in agreement with Fig. 6b. Stress drops were completely eliminated in meta-crystals containing 32 and 64 meta-grains (Fig. 7a). The hardening rates during plastic deformation of meta-crystals are shown in Fig. 7c. Compared to the large fluctuations of rate change in single meta-crystal, the hardening rates became more stable during deformation with a reduction in meta-grain size. Most notably, the stable and increasing hardening rates of all polygrain-like meta-crystals led to extended uniform plastic deformation with continuous increase in the flow stress. Flow stress at a given strain (e.g. $\sigma_{\varepsilon=0.3}$ at $\varepsilon = 0.3$) was quantified and plotted against $\frac{1}{\sqrt{d}}$ to demonstrate the effect of meta-grain boundaries on the strain hardening of meta-crystals. It was found that $\sigma_{\varepsilon=0.3} = 8 + \frac{43}{\sqrt{d}}$ well described the link between $\sigma_{\varepsilon=0.3}$ and the size of meta-grain. $k = 43$ higher than ratio 14.46 for normalised yield strength suggests that meta-grains boundaries were even more influential in plastic deformation (Fig. 7b).

The propagation and spatial distribution of shear bands were studied using DIC analyses to understand the role of meta-grain boundaries on the behaviours of meta-crystals during plastic deformation. Fig. 8 shows the shear bands distribution in meta-crystals at 20% strain. CAD models with the frame skin removed are shown to reveal the internal lattice orientation and struts connections across meta-grain boundaries. It shows that the shearing deformation in each meta-grain was dominated by lattice orientation similar to slip behaviour in crystals. A shear band system ($\langle 0\ 1\ 1 \rangle$ directions on $\{2\ 0\ 0\}$ planes) was observed in each meta-grain, which was consistent with that in the FCC45 single meta-crystal (Fig. 5c). The direction of shear band was deflected by around 90° across the meta-grain boundary due to the change in lattice orientations across the boundary. The decrease in the size of meta-grains significantly shortened the length of shear bands (Fig. 8a-e) in consistent with the observation of the twinned bi-meta-crystal (Fig. 6e), reaffirming the effect of meta-grain boundaries in restricting the propagation of shear bands. In addition, the distribution of shear bands is substantially more homogeneous throughout the meta-crystals with an increasing number of meta-grains (e.g. Fig. 8e vs a), making the deformation much more stable, explaining why the stress drop is increasingly diminished and ultimately eliminated with increasing the number of meta-grains as seen in Fig. 7. Moreover, meta-grain boundaries also temporarily stop the propagation of shear bands, limiting the transmission of shear band across boundaries; hence strengthening the meta-crystal during plastic deformation. The higher the density of boundaries (i.e. smaller size of meta-grains) the higher the induced hardening. Consequently, hardening rate increases with reducing the size of meta-grains (Fig. 7b and c). It is worth noting that the yield strength was measured as the stress at 0.2% strain offset. At this strain offset, the first shear band already occurred and experienced the stopping effect induced by meta-grain boundaries. This stopping effect therefore contributed to the yield strength when measured at the 0.2% strain offset, explaining the increase in measured yield stress with reducing the size of meta-grains as seen in Fig. 6 and 7.

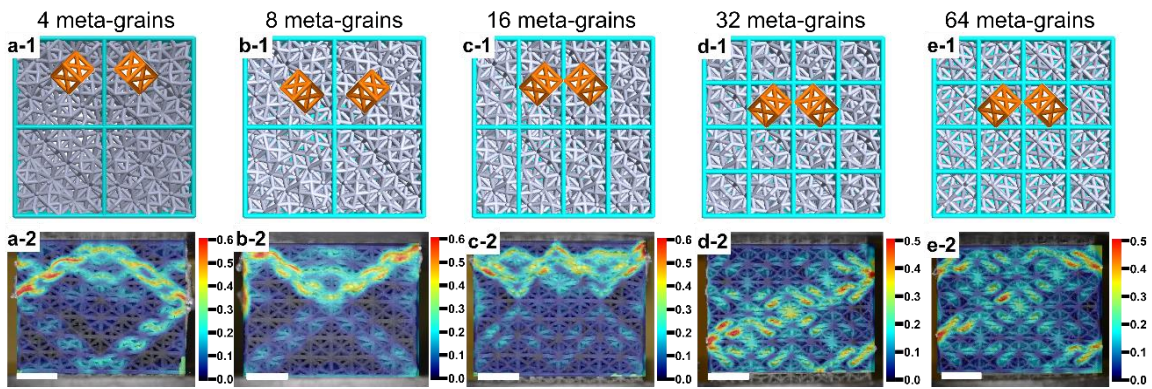

Fig. 8. Strain distribution at 20% strain on meta-crystals with respect to different lattice configurations after digital image correlation analyses (unit cell orientation and boundary location are highlighted), (a) 4 meta-grains, (b) 8 meta-grains, (c) 16 meta-grains, (d) 32 meta-grains, (e) 64 meta-grains, where the colour bar indicates the maximum shear strain and length scale bar is 10mm.

## 3.3. Boundary coherency: Effect of strut connectivity at boundaries

For polygrain alloys, it is suggested that the coherency of lattice at grain boundaries is one of the governing factors in the slip transmission, and hence the strengthening induced by grain boundaries [25,26]. Although the nature of bonding in crystals and meta-crystals is different (namely the bonding between atom-like nodes is made by physical struts while the connection in intrinsic crystals is by atomic bonds), the coherency of the lattice at boundaries is expected to be influential in meta-crystals. In particular, the change in lattice orientation across a meta-grain boundary disrupts the continuity of strut arrangements and might cause open-ended struts if not being deliberately connected by computer design. This might significantly weaken the meta-crystal of multiple meta-grains. To minimise such detrimental effect, 2D frames were introduced at a boundary to ensure the connectivity of struts to meta-grain boundaries (see Fig. 2c). Although the coherency at meta-grain boundaries is better represented by coincident site lattice as defined for intrinsic crystals [25,26], the connectivity of struts to an architected boundary also reflects the degree of coherency of architected polygrain-like materials. Fig. 9a shows the total number of struts connected to boundaries. It can be seen that the number of struts increased with decreasing the meta-grain size, which was mainly because of the increasing boundary areas. With the comparison to the measured yield stress for each meta-crystal shown in Fig. 9b, it is clear that the increasing trend of the yield stress (Fig. 9b) followed a similar trend of the total number of connected struts (Fig. 9a). Note that a meta-grain is comprised of complete unit cells (possessing same mechanical property and contribution to the global strength of meta-crystal) and open-ended struts connected to the boundaries. Therefore, higher connectivity of struts to boundaries made the boundaries better supported (i.e. stronger boundaries), resulting higher yield strength and effectively impeding shearing propagation.

The boundary connectivity depends on the types of boundary (namely tilt and twist types). According to the lattice orientation relationship used in the design of meta-crystals in this study, boundaries in the presented meta-crystals can be classified into three groups: group-1 (parallel to Y-Z plane), group-2 (parallel to X-Y plane) and group-3 (parallel to X-Z plane), where group-1 and -2 were of the tilt type and group-3 was of the twist type (Fig. 9c). Fig. 9d shows the percentage of the number of struts belong to each boundary group in relation to the total struts connecting to boundaries in the whole meta-crystal. Correlating Fig. 9b and d suggests that the introduction of tilt boundaries might induce more hardening effects on yield strength (e.g. 4 meta-grains and 32 meta-grains in Fig. 9b) than that of the twist boundary (e.g. 8 meta-grains and 64 meta-grains of which the fractions of twist boundary are relatively high). This is believed due to the connectivity of struts to different boundary types. The designed tilt boundary had higher coincident lattice sites (i.e. high coherency) at boundaries than the used twist boundary. Therefore, tilt boundaries were more supported and stronger, explaining higher extents of hardening in the tilt boundaries compared to the twist boundaries. In addition, the alignment of the tilt boundaries in the considered meta-crystals increased the interaction between the tilt boundaries with shear bands of {2 0 0}⟨0 1 1⟩, making the tilt boundaries more effective in stopping the transmission of shear bands across boundaries, hence more hardening induced by the tilt boundaries.

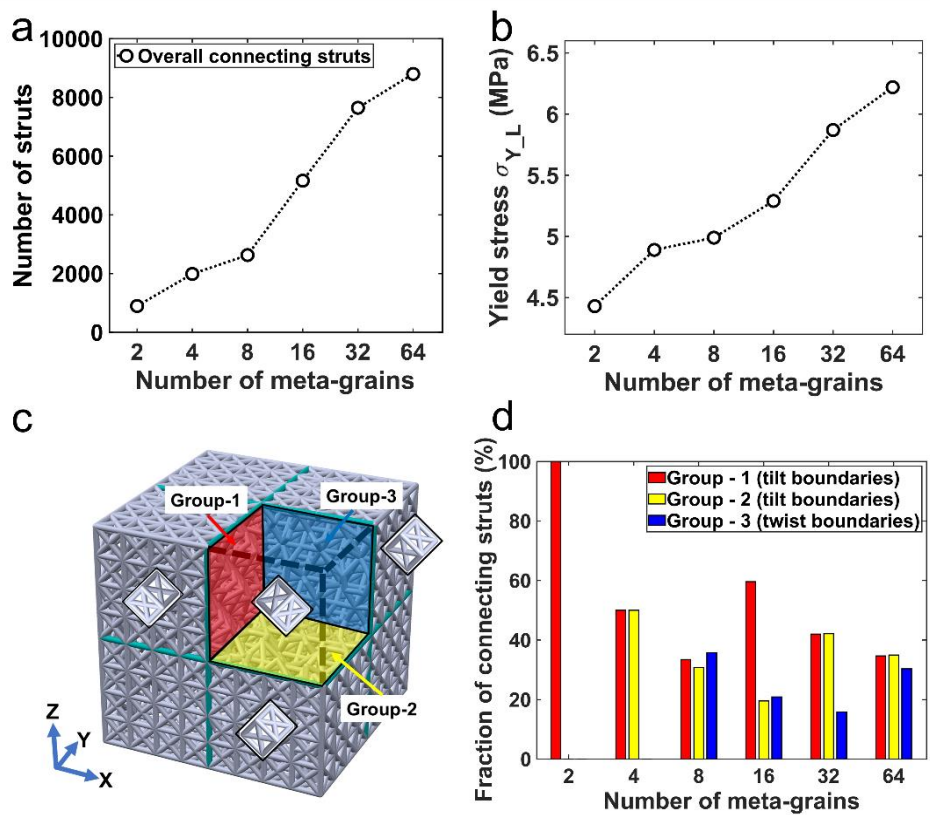

Fig. 9. (a) The total number of struts connecting to the meta-grain boundaries, (b) measured yield stress for meta-crystals containing different number of meta-grains, (c) schematic diagram illustrating the classification of boundaries with insets showing lattice orientation in each grain, (d) the fraction of struts connected to boundaries of a boundary type.

## 4. Conclusion

This study investigated the underlying mechanisms responsible for the shear band behaviour and hardening in architected polygrain-like materials (i.e meta-crystals). Buckling of individual struts was analysed to explain the occurrence of shear bands in two identified systems, i.e. $\{0\,2\,2\}\langle 1\,0\,0\rangle$ and $\{2\,0\,0\}\langle 0\,1\,1\rangle$ of FCC meta-crystals. The obtained understanding (including the identification of shear band systems) provides high confidence in predicting and controlling the shear bands and damage in polygrain-like meta-crystals.

Focus was made to reveal the underlying mechanisms responsible for the strengthening of meta-crystals containing polygrain-like domains (i.e. meta-grains). Specifically, the enhanced stability and strength of meta-crystals during plastic deformation is mainly due to the strength of meta-grain boundary that is associated with the boundary coherency. Similar to the role of coherency at grain boundaries in polycrystalline alloys, higher coherency results in better supported struts in architected polygrain-like meta-crystals, inducing higher yield strength. In particular, a higher number of total struts connecting to boundaries leads to higher strength. In addition to the influential role of boundary coherency, the effectiveness in changing the propagation of shear bands thanks to the lattice misorientation across meta-grain boundaries. Both the strengthened meta-grain boundaries and the misorientation make the boundaries obstacles to the propagation of shear bands. Meta-grain boundaries can stop, shorten and deflect shear bands during post-yielding deformation. It is worth noting that the yield strength was measured at strain offset of 0.2% at which the first shear band already happens and experiences the stopping effect of meta-grain boundaries on the shear band propagation, explaining the increase in macroscopically measured yield strength with decreasing the meta-grain size. An architected tilt type of boundaries induced higher hardening than an architected twist one, highlighting that different coherency of meta-grain boundaries results in different degrees of hardening. This implies that high strength architected materials can be designed without reducing the size of meta-grains, but via engineering the coherency similar to the grain engineering approach in metallurgy.

**Declaration of Competing Interest**

The authors declare that there is no known competing interest.

**Acknowledgements**

The authors thank financial support via the Excellent Funds for Frontier Research provided by Imperial College London.

**Author contributions**

**Chen Liu:** Conceptualisation, Methodology, Software, Formal analysis, Investigation, Data curation, Writing – Original draft, Writing – Review & Editing. **Jedsada Lertthanasarn:** Formal analysis, Investigation, Writing – Review & Editing. **Minh-Son Pham:** Conceptualisation, Resources, Formal analysis, Supervision, Funding acquisition, Writing – Review & Editing.

experimental test and μ-CT-based finite element analysis, Mater. Des. (2019). https://doi.org/10.1016/j.matdes.2019.107685.

[29] H. Wang, Y. Fu, M. Su, H. Hao, Effect of structure design on compressive properties and energy absorption behavior of ordered porous aluminum prepared by rapid casting, Mater. Des. (2019). https://doi.org/10.1016/j.matdes.2019.107631.